\documentclass[rnaas]{aastex62}
\usepackage{amsmath}

\begin{document}

\title{Classifying Signatures of Sudden Ionospheric Disturbances}

\author[0000-0002-9370-8061]{Sahil Hegde}
\affiliation{Department of Physics, Columbia University, New York, NY 10027, USA}

\author[0000-0002-5662-9604]{Monica G. Bobra}
\affiliation{W. W. Hansen Experimental Physics Laboratory, Stanford University, Stanford, CA 94305, USA}

\author[0000-0002-6937-6968]{Philip H. Scherrer}
\affiliation{W. W. Hansen Experimental Physics Laboratory, Stanford University, Stanford, CA 94305, USA}

\correspondingauthor{Monica Bobra}
\email{mbobra@stanford.edu}


\begin{abstract}
Solar activity, such as flares, produce bursts of high-energy radiation that temporarily enhance the D-region of the ionosphere and attenuate low-frequency radio waves. To track these Sudden Ionospheric Disturbances (SIDs), which disrupt communication signals and perturb satellite orbits, \citet{Scherrer2008} developed an international, ground-based network of around 500 SID monitors that measure the signal strength of low-frequency radio waves. However, these monitors suffer from a host of noise contamination issues that preclude their use for rigorous scientific analysis. As such, we attempt to create an algorithm to automatically identify noisy, contaminated SID data sets from clean ones. To do so, we develop a set of features to characterize times series measurements from SID monitors and use these features, along with a binary classifer called a support vector machine, to automatically assess the quality of the SID data. We compute the True Skill Score, a metric that measures the performance of our classifier, and find that it is $\sim$0.75$\pm$0.06. We find features characterizing the difference between the daytime and nighttime signal strength of low-frequency radio waves most effectively discern noisy data sets from clean ones. 

\end{abstract}

\section{Introduction} 

The Earth's ionosphere exhibits diurnal changes in composition and structure due to solar radiation. In the daytime, incident solar photons ionize the thin, lowest-lying layer of the ionosphere called the D-region. During the night, the D-region effectively disappears. Solar activity, such as flares, produce bursts of Extreme Ultraviolet (EUV) and X-Ray radiation that temporarily enhance the D-region. These enhancements are known as Sudden Ionospheric Disturbances (SIDs) and can disrupt communication signals and affect atmospheric drag. During a SID, the ion density in the D-region attenuates radio waves. 

Since this attenuation varies as the inverse square of the frequency, \citet{Scherrer2008} developed an international, ground-based network of around 500 SID monitors that measure the signal strength of extremely low frequency (ELF) and very low frequency (VLF) radio waves from shore radio stations, which transmit one-way communication to submarines, to probe the structure of the ionosphere (see \url{sid.stanford.edu}). SID monitors measure ELF or VLF signal strength continuously, at a cadence of once every 5 seconds. However, these monitors suffer from noise contamination issues that preclude their use for rigorous scientific analysis. As such, we developed an algorithm to automatically identify noisy, contaminated time series measurements from clean ones. 

\section{Method}

To do this, we first manually label 200 24-hour time series measurements of ELF and VLF signal strength, as measured by the SID monitors, as clean data (or members of the positive class) or noisy data (members of the negative one). We then calculate five features on each of these time series measurements:

\begin{enumerate}

\item The difference in average signal strength ($\phi$) between the day and night segments:
$$\textup{F}1 = \bar{\phi}_{night} - \bar{\phi}_{day}.$$

\item  The similarity between a baseline daytime time series and any given daytime time series using Dynamic Time Warping (DTW; \citealt{Salvador2007}).

\item The similarity between a baseline sunrise signature, manifested as a sudden decrease in signal strength, and any given time series using the same DTW technique.

\item The difference in the variance of signal strength the between day and night segments:
$$\textup{F}4 = \operatorname {Var} (\phi_{night}) - \operatorname {Var} (\phi_{day}).$$

\item To quantify the signal$-$to$-$noise in any given time series, we compute the DTW distance between a smoothed baseline time series and a smoothed version of any given time series. 

\end{enumerate}

\begin{figure}
    \centering
    \includegraphics[scale=1.1]{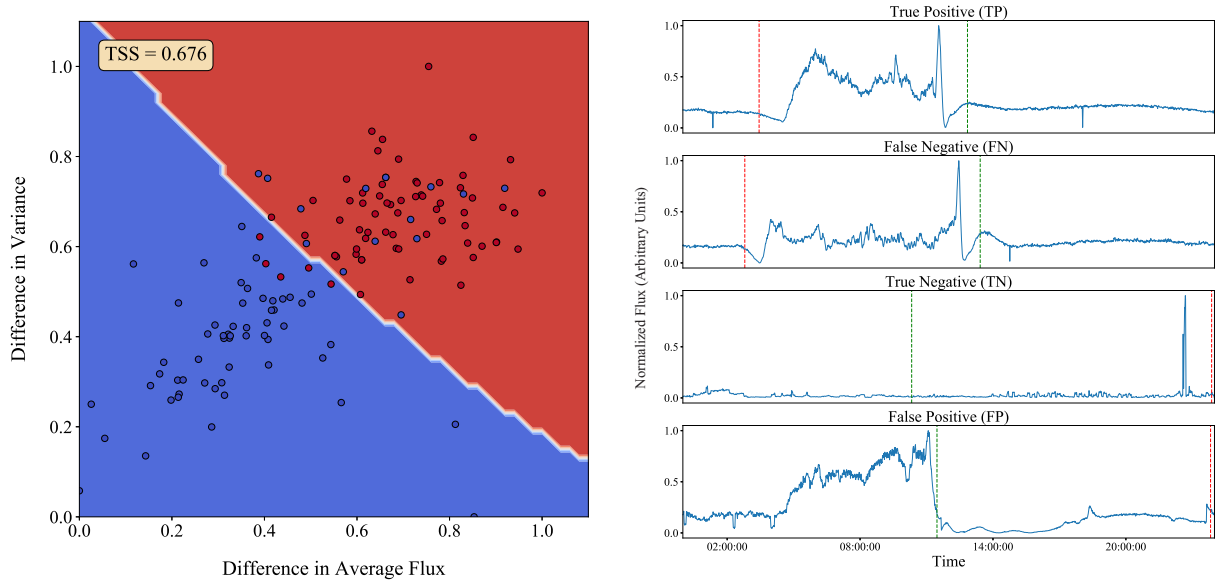}
    \caption{The left panel shows a visual representation of the decision boundary, computed by the SVM, for our two most effective features --- features 1 and 4. The positive class is indicated in red, and the negative in blue. We see more misclassifcations for negative examples than positive ones. Further, misclassified positive examples are fairly close to the decision boundary. On the right, we display an example time series for each possible classification. Since SID monitors measure different radio frequencies across the ELF and VLF band, we normalize the signal strength across all monitors and plot these time series with those normalized units. The green dashed lines indicates sunrise, and the red ones indicate sunset. True positives were highly dependent on average signal strength (feature 1) and variance (feature 4). False positives occurred when noisy data yielded similar values for features 1 and 4. True negatives showed little signal throughout the 24-hour period. False negatives usually had a lower nighttime signal strength.}

    \label{fig:figurename}
\end{figure}

Finally, we use a binary classifer, called a Support Vector Machine (SVM; \citealt{Pedregosa2011}), to classify the data. The SVM constructs an n-dimensional feature space (in our case, n=5), plots all the features, and draws a non-linear decision boundary that separates the features associated with the positive class from the features associated with the negative ones. For more detail, see \citet{Bobra15}.

We train the SVM on 70\% of our data and test it on the remainder. We evaluate performance of the SVM with a binary classification metric called the True Skill Score \citep{bloomfield2012}: 
$$\textup{TSS} = \frac{\textup{TP}}{\textup{TP+FN}} - \frac{\textup{FP}}{\textup{FP+TN}}$$
This score ranges from [-1,1], with 0 representing no predictive power. We shuffle our data into 100 different training and testing sets, computing the TSS each time. We take the average of these 100 scores to obtain the final TSS, and the standard deviation of these scores to obtain the error in the TSS.

\section{Conclusion}

We find that our algorithm can effectively distinguish between clean and noisy data, yielding a TSS of 0.757 $\pm$ 0.068 using all five features. We also find that features 1 and 4, the difference in flux and variance, most effectively classify our data. We surmise that a larger training set, a better curve matching algorithm, and more representative features for data noise could improve the classifier's performance.

The code and data we used to conduct this analysis are publicly available at the Stanford Digital Repository (\url{https://purl.stanford.edu/cs332mr4558}).

\acknowledgments

We gratefully acknowledge the software used to make this work possible: \texttt{scikit-learn} \citep{Pedregosa2011}, \texttt{astropy} \citep{Astropy2013, Astropy2018}, \texttt{sunpy} \citep{SunPy2015}, \texttt{FastDTW} \citep{Salvador2007}, \texttt{TensorFlow} \citep{tensorflow2015}, and all the packages that support these ones.

\bibliography{bibliography.bib}

\end{document}